\documentclass[twocolumn,prd,nofootinbib]{revtex4}

\usepackage{amsmath,amssymb,amsfonts}
\usepackage{mathrsfs}
\usepackage{graphicx}% Include  filesg
\usepackage{subfigure}
\usepackage[]{natbib}
\usepackage{enumerate}
\usepackage{color}

\newcommand{\Deqn}[1]{{Eq.~(\ref{#1})}}
\newcommand{\Deqns}[1]{{Eqs.~(\ref{#1})}}

\newcommand{\Dfig}[1]{{Fig.~\ref{#1}}}
\newcommand{\beq}{\begin{equation}}
\newcommand{\eeq}{\end{equation}}
\newcommand{\bea}{\begin{eqnarray}}
\newcommand{\eea}{\end{eqnarray}}
\newcommand{\tr}{{\mathop{\mathrm{tr}}\nolimits}}
\newcommand{\be}{\begin{equation}}
\newcommand{\ee}{\end{equation}}

\newcommand{\lp}{\left(}
\newcommand{\rp}{\right)}
\newcommand{\bb}{\begin{bmatrix}}
\newcommand{\eb}{\end{bmatrix}}

\begin{document}

\title{Time-domain Implementation of the Optimal Cross-Correlation 
Statistic for Stochastic Gravitational-Wave Background Searches 
in Pulsar Timing Data}

\author{Sydney J. Chamberlin} 
\email{sydc@gravity.phys.uwm.edu}
\affiliation{Center for Gravitation, Cosmology, and Astrophysics, Department of Physics, 
University of Wisconsin--Milwaukee, Milwaukee, Wisconsin 53201, 
USA}

\author{Jolien D. E. Creighton}
\email{jolien@gravity.phys.uwm.edu}
\affiliation{Center for Gravitation, Cosmology, and Astrophysics, Department of Physics, 
University of Wisconsin--Milwaukee, Milwaukee, Wisconsin 53201, 
USA}

\author{Paul Demorest}
\email{pdemores@nrao.edu}
\affiliation{National Radio Astronomy Observatory, Charlottesville, Virginia 22903 
USA}

\author{Justin Ellis}
\email{Justin.A.Ellis@jpl.nasa.gov}
\affiliation{Jet Propulsion Laboratory, California Institute of Technology, 
Pasadena, California 91109, USA}
\altaffiliation{Einstein Fellow}

\author{Larry R. Price} 
\email{larryp@caltech.edu}
\affiliation{LIGO Laboratory, California Institute of Technology, Pasadena, California 91125, 
USA}

\author{Joseph D. Romano} 
\email{joe@phys.utb.edu}
\affiliation{Department of Physics and Astronomy and 
Center for Gravitational-Wave Astronomy, 
University of Texas at Brownsville, Brownsville, Texas 78520, 
USA}

\author{Xavier Siemens}
\email{siemens@gravity.phys.uwm.edu}
\affiliation{Center for Gravitation, Cosmology, and Astrophysics, Department of Physics, 
University of Wisconsin--Milwaukee, Milwaukee, Wisconsin 53201, 
USA}

\date{\today}

\begin{abstract} 

Supermassive black hole
binaries, cosmic strings, relic gravitational waves from
inflation, and first order phase transitions in the early universe
are expected to contribute to a stochastic background of
gravitational waves in the $10^{-9}$~Hz--$10^{-7}$~Hz frequency
band.  Pulsar timing arrays (PTAs) exploit the high precision timing
of radio pulsars to detect signals at such frequencies. 
Here we present a time-domain implementation of the optimal
cross-correlation statistic for stochastic background searches in
PTA data. Due to the irregular sampling typical
of PTA data as well as the use of a timing model to predict
the times-of-arrival of radio pulses, time-domain methods are better
suited for gravitational wave data analysis of such data. We present a
derivation of the optimal cross-correlation statistic starting from
the likelihood function, a method to produce simulated stochastic
background signals, and a rigorous derivation of the scaling laws for
the signal-to-noise ratio of the cross-correlation statistic in
the two relevant PTA regimes: the weak signal limit where instrumental
noise dominates over the gravitational wave signal at all frequencies,
and a second regime where the gravitational wave signal dominates at
the lowest frequencies.    
\end{abstract}

\maketitle

\section{Introduction}

Gravitational waves, a key prediction of Einstein's theory of general
relativity, are perturbations in the fabric of spacetime produced by
the accelerated motion of massive objects. The direct detection of
gravitational waves is likely to occur in the next few years, and
promises to provide a new means to study the universe.  A number of
worldwide efforts aiming to detect gravitational waves are currently
underway.  At the low-frequency end of the detectable
gravitational-wave spectrum ($10^{-9}$ Hz--$10^{-7}$ Hz), pulsar
timing arrays (PTAs) exploit the remarkable high precision timing of
radio pulsars to search for gravitational waves
\cite{haa+10b}. Pulsars have already been used to indirectly measure
the effects of gravitational-wave emission through the Hulse-Taylor
binary~\cite{ht75}. A direct detection of gravitational waves is
possible with an array of precisely-timed pulsars: a gravitational
wave propagating through spacetime affects the travel time of radio
pulses from pulsars, and can be observed by searching for correlated
deviations in the expected times-of-arrival of the radio
pulses~\cite{saz78, det79}.

The most likely source of gravitational waves at nanoHz frequencies
are supermassive black hole binaries (SMBBHs) that form following the
merger of massive galaxies~\cite{jb03,svc08,svv09}.  The superposition
of gravitational waves from all SMBBH mergers forms a stochastic
background of gravitational
waves~\cite{lb01,jb03,wl03,vhm03,ein+04,svc08,s12,mop12}.  Individual
periodic signals~\cite{svv09,sv10,rs11,rwh+12,mgs+12} and
bursts~\cite{vl10,Cordes:2012zz} can also be produced by SMBBH
systems.  In addition, cosmic strings~\cite{oms10, sa13, mn13,
  kms+13}, first order phase transitions in the early
universe~\cite{ccd+10}, and relic gravitational waves from
inflation~\cite{s79, zzy+13} are potential sources of gravitational
waves in the nanoHertz band.

A number of data analysis techniques have been developed and
implemented to search for isotropic stochastic backgrounds of
gravitational waves in PTA  data~\cite{det79,srt+90,l02,jhl+05,jhs+06,abc+09,%
hlm+09,vlm+09,ych+11,vhj+11,cs12,dfg+12,%
rwh+12, bp12, vhl12, esvh13}.  More recently, these techniques have
been generalized to searches for anisotropic backgrounds~\cite{msm+13, tg13,grt+14,ms14}.  
Additionally, a range of data analysis methods have been developed 
to search for individual periodic sources that stand out over the
stochastic background~\cite{svv09, sv10, rs11, esc12, mgs+12,pbs+12,%
 ejm12, tyz+13, teg14, abb+14, wmj14, yss14, zhw+14},  
bursts~\cite{fl10, cj12, pit12, mcc14, deng14}, and signals of
unknown form~\cite{cbv+14}.

In this paper we describe a practical time-domain implementation of the optimal
cross-correlation statistic~\cite{abc+09} that can be used to search for 
isotropic stochastic backgrounds. 
In Section~\ref{sec:PTAoverview}, we review the effect of a
gravitational wave on the pulsar-Earth system, and the expected
cross-correlations in the times-of-arrival of pulses from different
pulsars. In Section~\ref{sec:optStat}, we develop the formalism needed
to implement the search for a stochastic background, including the
timing model, and derive the optimal cross-correlation statistic from
the likelihood ratio.  In Section~\ref{sec:inj}, we develop a
procedure for injecting simulated stochastic background signals into
PTA data, and in Section~\ref{sec:scalingLaws}, we describe the
scaling laws that govern the expected signal-to-noise ratio of the
cross-correlation statistic.  We conclude in
Section~\ref{sec:conclude} with a discussion of the practicality of
implementing the statistics introduced in this paper for
gravitational-wave searches.  For reference, we will work in units
where $c = G = 1$.

\section{Preliminaries}
\label{sec:PTAoverview}

An array of pulsars can be used to search for a stochastic background
of gravitational waves. Deviations from the expected times-of-arrival
of pulses from different pulsars are correlated, and with enough
timing precision these correlations are measurable. In this section we
describe how the times-of-arrival of pulses from pulsars are affected
by gravitational waves, and discuss the expected correlation of
signals from different pulsars.
%%%%%% Pulsar-Earth figure %%%%%%%%%
\begin{figure}
\includegraphics[width=3.4in]{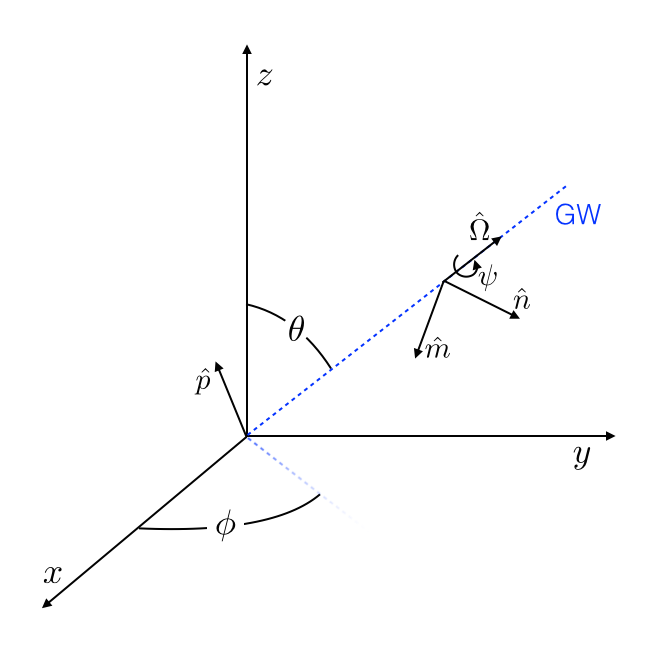}
\caption{(color online) The pulsar-Earth system, as visualized with
  the Earth at the origin. The gravitational wave propagates as the
  blue dashed line, and the vectors defined in
  \Deqns{omega},~(\ref{m}) and~(\ref{n}) are included with polar and
  azimuthal angles. The angle $\psi$ designates the polarization angle
  of the gravitational wave. For a stochastic gravitational wave
  background, this angle is averaged over many independent sources
  and can be chosen to be zero.}
\label{position_diagram}
\end{figure}
%%%%%%%%%%%%%%%%%%%%%%%%%

Gravitational waves induce a redshift in the signal from the pulsar
that depends on the geometry of the pulsar-Earth system and the metric
perturbation~\cite{det79}.  For a pulsar located in the direction of
unit vector $\hat{p}$ (that points from Earth to the pulsar), and a
gravitational wave propagating in the direction $\hat{\Omega}$ (see
\Dfig{position_diagram}), the redshift induced in the radio pulse is
proportional to the change in the metric perturbation at the Earth,
when the pulse is received, and at the pulsar, when the pulse is
emitted~\cite{det79,abc+09}
\beq
z(t,\hat{\Omega}) = 
\frac{1}{2}
\frac{\hat{p}^i\hat{p}^j}{1+\hat{\Omega}\cdot\hat{p}}\Delta h_{ij},
\label{tdredshift}
\eeq
where\footnote{Here we correct
a sign error in a previous paper~\cite{abc+09}, pointed out to us by
Eanna Flanagan.}
\beq
\Delta h_{ij}
\equiv
h_{ij}(t_{\rm e},\hat{\Omega}) - 
h_{ij}(t_{\rm p},\hat{\Omega})
\label{delhdef}
\eeq and $i,j$ denote spatial components\footnote{Note that in this
  section we use the Einstein summation notation where repeated
  indices are summed over.}.  These terms are typically referred to as
the Earth-term and the pulsar-term, respectively.

The total redshift is obtained by integrating \Deqn{tdredshift} over all directions on the sky
\beq
z(t) = \int_{S^2} d\hat{\Omega} \, z(t,\hat{\Omega}).
\label{zftot}
\eeq
It is important to point out that in pulsar timing the observable quantity is 
actually not the redshift, but the \emph{timing residual}, which is
just the integral of the redshift
\beq
r(t) = \int_0^{t} dt' \, z(t').
\label{residual}
\eeq

The metric perturbation in terms of the 
usual plane wave expansion is~\cite{ar99}
\beq
h_{ij}(t,\vec{x}) = \sum_{A}
\int_{-\infty}^{\infty}df\, \int_{S^2}d\hat{\Omega}\, 
e^{i2\pi f(t-\hat{\Omega}\cdot\vec{x})}
h_A(f,\hat{\Omega})e_{ij}^A(\hat{\Omega}),
\label{pwexp}
\eeq
where $f$ is the frequency of the gravitational wave, $A = +, \times$
labels the polarization modes, and $e_{ij}^A(\hat{\Omega})$ are the
polarization tensors (see below). We can use this expansion to write a
frequency-domain expression for the timing residuals produced by a
gravitational wave traveling in the direction $\hat
\Omega$. Specifically,
\begin{multline}
\tilde{r}(f,\hat{\Omega}) 
= \frac{1}{2 \pi i f}\left(1-e^{-2\pi i f L(1+\hat{\Omega}\cdot\hat{p})}\right) 
\\
\times \sum_{A} h_A(f,\hat{\Omega}) \left ( e^A_{ij}(\hat{\Omega})
\frac{\hat{p}^i\hat{p}^j}{2(1+\hat{\Omega}\cdot\hat{p})}\right),
\label{fdredshift}
\end{multline}
where $L$ is the pulsar-Earth distance.

The polarization tensors are
 \begin{subequations}
\begin{align}
\label{e_plus}
  e_{ij}^+({\hat{\Omega}}) &=  {\hat{m}}_i {\hat{m}}_j - {\hat{n}}_i {\hat{n}}_j,\\
\label{e_cross}
  e_{ij}^{\times}({\hat{\Omega}}) &= {\hat{m}}_i {\hat{n}}_j + {\hat{n}}_i {\hat{m}}_j,
\end{align}
\end{subequations}
 where the quantities 
\begin{subequations}
\begin{align}
\label{omega}
  {\hat{\Omega}}&= (\sin{\theta} \cos{\phi},  \sin{\theta} \sin{\phi},  
\cos{\theta})=\hat r, \\
\label{m}
  {\hat{m}}&=(\sin{\phi}, -\cos{\phi}, 0)=-\hat\phi, \\
\label{n}
  {\hat{n}}&=(\cos{\theta}\cos{\phi}, \cos{\theta}\sin{\phi}, -\sin{\theta})=\hat\theta,
\end{align}
\end{subequations}
describe the geometry of the propagating gravitational wave as shown in \Dfig{position_diagram}.

The energy density of gravitational waves is given by
\beq
\rho_{\rm gw} = \frac{1}{32\pi}\langle \dot{h}_{ij}(t,\vec x)
\, \dot{h}^{ij}(t,\vec x)\rangle,
\label{eqrho}
\eeq
and the spectrum of a stochastic background is  
\beq
 \Omega_{\rm gw}(f)\equiv \frac{1}{\rho_{\rm crit}}\frac{d\rho_{\rm gw}}{d\ln f},
\label{omdef}
\eeq
where $\rho_{\rm crit} = 3 H_0^2/(8\pi)$ is the critical energy density,
and $H_0$ is the Hubble constant. 

If we assume the stochastic gravitational wave background  is isotropic, unpolarized, and stationary 
the correlation function for the strain can be written as~\cite{ar99,abc+09}
\begin{multline}
\langle h_A^*(f,\hat{\Omega})h_{A'}(f',\hat{\Omega}')\rangle 
=\frac{3H_0^2}{32\pi^3}\delta^2(\hat{\Omega},\hat{\Omega}')\delta_{AA'}
\delta(f-f')
\\
\times|f|^{-3}\Omega_{\rm gw}(|f|),
\label{hevom}
\end{multline}
and therefore the expected correlation in the timing residuals is
\beq\label{e:rexpect}
\langle\tilde{r}_I^*(f)\tilde{r}_J(f')\rangle
= \frac{H_0^2}{16 \pi^4} \delta(f-f')|f|^{-5}
\Omega_{\rm gw}(|f|) \chi_{IJ},
\eeq
where $\chi_{IJ}$ is the Hellings and Downs
coefficient for pulsars $I$ and $J$~\cite{hd83}
 \be 
\chi_{IJ}=\frac{3}{2} \left[ \frac{1}{3} +
  \frac{1-\cos\zeta_{IJ}}{2}\left[ \ln\lp\frac{1-\cos\zeta_{IJ}}{2}\rp
    - \frac{1}{6} \right] \right] +\frac{1}{2}\delta_{IJ},
\label{hdcurve}
\ee  
and $\zeta_{IJ}$ is the angular separation of the pulsar
pair. 

In parts of this paper, we will refer not to
$\Omega_{\rm gw}(f)$ but
instead to the dimensionless gravitational wave amplitude $A_{\rm gw}$
(at reference frequency $f_{\rm 1yr}={\rm yr}^{-1}$) which
appears in the expression for the characteristic strain
\be 
h_c(f) = A_{\rm gw} \left(\frac{f}{f_{\rm 1yr}} \right)^\alpha.
\ee
The spectral index $\alpha$ depends on the astrophysical source of the
background. For example, a stochastic background produced by supermassive black hole binary
systems has $\alpha = -2/3$ \cite{jb03, svc08}.
The amplitude $A_{\rm gw}$ is related to the strain
spectral density $S_h(f)$ of the gravitational-wave 
background via:
\be
S_h(f) = \frac{h_c^2(f)}{f}.
\ee
For one-sided power spectra, $S_h(f)$ and $A_{\rm gw}$
are related to $\Omega_{\rm gw}(f)$ by
\begin{align}
S_h(f) &=\frac{3H_0^2}{2\pi^2}\frac{\Omega_{\rm gw}(f)}{f^3},
\\
\Omega_{\rm gw}(f) &= \frac{2\pi^2}{3 H_0^2} 
A_{\rm gw}^2 f^2 \left(\frac{f}{f_{\rm 1yr}}\right)^{2 \alpha}. 
\label{omegatoA} 
\end{align}
Note that in this paper we will work exclusively with one-sided
spectra, which differs from the convention adopted in
\cite{sej+13}.

\section{The optimal cross-correlation statistic}
\label{sec:optStat}

\subsection{Timing Model}
\label{sec:timingModel}

In pulsar timing experiments the quantities that are directly measured
are the times-of-arrival (TOAs) of radio pulses emitted from
pulsars. These TOAs contain many terms of known functional form,
including intrinsic pulsar parameters (pulsar period, spin-down,
etc.), along with stochastic processes such as radiometer noise, pulse
phase jitter, and possibly red noise either from interstellar medium
(ISM) effects, intrinsic pulsar noise, and, potentially, a
gravitational wave background.

Suppose that the TOAs for a pulsar are given by
\be
\mathbf{t}^{\rm obs}=\mathbf{t}^{\rm det}(\boldsymbol\xi_{\rm true})+\mathbf{n},
\ee
where $\mathbf{t}^{\rm obs}$ are the $N_{\rm TOA}$ observed TOAs, $\mathbf{t}^{\rm det}$ are the deterministic modeled TOAs parameterized by  $N_{\rm par}$ timing model parameters $\boldsymbol\xi_{\rm true}$, and $\mathbf{n}$ is the noise time series in the measurement which is assumed to be Gaussian with covariance matrix given by
\be
\mathbf{N}=\langle \mathbf{n}\mathbf{n}^{T} \rangle=\mathbf{N}^{\rm white}+\mathbf{N}^{\rm red}
\label{covy}
\ee
where the $N_{\rm TOA} \times N_{\rm TOA}$ matrtices $\mathbf{N}^{\rm
  white}$ and $\mathbf{N}^{\rm red}$ are the contributions to the
covariance matrix from the white and red noise processes, respectively. 
We will discuss the exact form of this covariance matrix in the next section. 
Assuming that estimates of the true timing model parameters 
$\boldsymbol\xi_{\rm est}$ exist (either from information gained when 
discovering the pulsar or from past timing observations), 
we can form the pre-fit timing residuals as
\be
\delta \mathbf{t}^{\rm pre}=\mathbf{t}^{\rm obs}-\mathbf{t}^{\rm det}(\boldsymbol\xi_{\rm est})=\mathbf{t}^{\rm det}(\boldsymbol\xi_{\rm true})+\mathbf{n}-\mathbf{t}^{\rm det}(\boldsymbol\xi_{\rm est}).
\ee
As mentioned above, we will assume that the initial estimates 
for our timing model parameters are correct to some linear offset 
$\boldsymbol\xi_{\rm est}=\boldsymbol\xi_{\rm true}+\delta\boldsymbol\xi$, 
for which the pre-fit residuals become
\be
\delta \mathbf{t}^{\rm pre}=
\mathbf{t}^{\rm det}(\boldsymbol\xi_{\rm true})-\mathbf{t}^{\rm det}(\boldsymbol\xi_{\rm true}+\delta\boldsymbol\xi)+\mathbf{n}.
\ee
Expanding this solution around the true timing model parameters, we obtain
\be
\label{pre}
\begin{split}
\delta \mathbf{t}^{\rm pre}&=\left.
-\frac{\partial \mathbf{t}^{\rm det}}{\partial\boldsymbol\xi}\right|_{\boldsymbol\xi=\boldsymbol\xi_{\rm true}}
\delta\boldsymbol\xi+\mathbf{n}+\mathcal{O}(\delta\boldsymbol\xi^{2})\\
&\approx \left.-\frac{\partial \mathbf{t}^{\rm det}}{\partial \boldsymbol\xi}\right|_{\boldsymbol\xi=\boldsymbol\xi_{\rm true}}\delta\boldsymbol\xi+\mathbf{n}\\
&= \mathbf{M}\delta\boldsymbol\xi+\mathbf{n},
\end{split}
\ee
where $\mathbf{M}$ is an $N_{\rm TOA} \times N_{\rm par}$ matrix, commonly referred to as the \emph{design matrix} \cite{pvt+92, d07}. Here we have assumed that our initial estimate of the model parameters is sufficiently close to the true values so that we can approximate this as a linear system of equations in $\delta\boldsymbol\xi$. It is customary in standard pulsar timing analysis to obtain the best fit $\delta\boldsymbol\xi$ values through a generalized least-squares minimization of the pre-fit residuals.  The function that we seek to minimize is (see \cite{hem06})
\be
\chi^{2}=\frac{1}{2}(\delta\mathbf{t}^{\rm pre}-\mathbf{M}\delta\boldsymbol{\xi})^T \mathbf{N}^{-1}(\delta\mathbf{t}^{\rm pre}-\mathbf{M}\delta\boldsymbol{\xi}).
\ee
Minimizing this function with respect to the parameter offsets $\delta\boldsymbol\xi$ results in
\be
\delta\boldsymbol\xi_{\rm best}=-\lp \mathbf{M}^{T}\mathbf{N}^{-1}\mathbf{M} \rp^{-1}\mathbf{M}^{T}\mathbf{N^{-1}}\delta\mathbf{t}^{\rm pre}.
\ee
The post-fit residuals are then given by
\be
\delta\mathbf{t}^{\rm post}\equiv
\delta \mathbf{t}^{\rm pre}-\mathbf{M}\delta\boldsymbol\xi_{\rm best} 
= \mathbf{R}\delta\mathbf{t}^{\rm pre},
\label{resids}
\ee
where 
\be
\mathbf{R}=\mathbb{I}-\mathbf{M}\lp \mathbf{M}^{T}\mathbf{N}^{-1}\mathbf{M} \rp^{-1}\mathbf{M}^{T}\mathbf{N}^{-1}
\label{Rmatrices}
\ee
is a an $N_{\rm TOA} \times N_{\rm TOA}$ oblique projection matrix
that transforms pre-fit to post-fit residuals, and $\mathbb{I}$ is the
identity matrix. All of the information about any noise source or
stochastic gravitational-wave background is encoded in
$\mathbf{N}$. However, in most cases we have no a priori knowledge of
this covariance matrix and therefore assume that it is given by
$\mathbf{W}=\rm{diag}(\{\sigma_{i}^{2}\})$, where $\sigma_{i}$ is the
uncertainty of the $i$th TOA. Previous work \citep{chc+10} has used an
iterative method to estimate the covariance matrix of the residuals
and apply a generalized least squares fit.  For this work we will only
work with residuals that have been created using a weighted least
squares fit.  It should be noted that in standard pulsar timing
packages such as \textsc{tempo2} \cite{hem06} this process must be
iterated. In other words, the pre-fit residuals are formed with an
initial guess of the parameters, and the chi-squared is then minimized
to produce best estimates of the parameters. This may not be a good
fit, however, as we have assumed that the pre-fit residuals are linear
in the parameter offsets. Consequently, we form new parameter
estimates from the best fit parameter offsets and iterate until the
fit converges, with the reduced chi-squared serving as the
goodness-of-fit parameter. For this reason, we must ensure that our
timing model fit has converged prior to any gravitational-wave
analysis.

\subsection{Derivation of the optimal statistic}
\label{sec:derOptStat}

\subsubsection{Likelihood function for a PTA}

Much of the discussion in this section follows closely that
of~\cite{esvh13}, with some of the details
included here. We begin by assuming that our PTA consists of $M$
pulsars, each with some intrinsic noise $\mathbf{n}_I(t)$. Henceforth
uppercase latin indices will label a pulsar and lowercase latin
indices will label a particular TOA. Under the assumption that all
intrinsic pulsar noise is Gaussian, we can write the full likelihood
function for the PTA as
\be
\label{likey}
p(\mathbf{n}|\vec\theta)=\frac{1}{\sqrt{\det(2 \pi \mathbf{\Sigma}_{n})}}\exp\lp-\frac{1}{2}\mathbf{n}^T\mathbf{\Sigma}_{n}^{-1}\mathbf{n}\rp,
\ee
where now we are using the full PTA noise time series that is just a concatenated length $M N_{\rm TOA}$ column vector 
\be
\mathbf{n}=\bb \mathbf{n}_{1} \\ \mathbf{n}_{2}\\ \vdots \\ \mathbf{n}_{M} \eb,
\ee
$\mathbf{\Sigma}_{n}$ is the $M N_{\rm TOA} \times M N_{\rm TOA}$
covariance matrix and  $\vec\theta$ is a set of
parameters that characterize the noise. 
 The covariance matrix for the noise is the block matrix
\be \label{cov} \boldsymbol{\Sigma}_n=\bb \mathbf{N}_{1} & \mathbf{X}_{12} & \hdots &
\mathbf{X}_{1M}\\ \mathbf{X}_{21} & \mathbf{N}_{2} & \hdots & \mathbf{X}_{2M}\\ \vdots & \vdots & \ddots &
\vdots\\ \mathbf{X}_{M1} & \mathbf{X}_{M2} & \hdots & \mathbf{N}_{M}\eb, \ee 
where 
\begin{align}
\mathbf{N}_{{I}}&=\langle \mathbf{n}_{I}\mathbf{n}_{I}^{T}\rangle,\\
\mathbf{X}_{IJ}&=\langle \mathbf{n}_{I}\mathbf{n}_{J}^{T}\rangle\big|_{I\ne J}, 
\end{align} 
are the auto-covariance and cross-covariance matrices, respectively, for each set of noise vectors. 

In general the autocorrelation matrices are defined via the  
Wiener--Khinchin theorem  as
\be
\begin{split}
\mathbf{N}_I=\langle \mathbf{n}_I \mathbf{n}_I^T \rangle_{ij} 
&=  \int_0^{\infty} df e^{2\pi i f \tau_{ij}}  \mathcal{P}_I(f)\\
&+  \mathcal{F}_I\mathbf{W}_I+\mathcal{Q}_I^2\mathbb{I} 
\end{split}
\ee
where $\tau_{ij}=|t_i-t_j|$,  $\mathcal{F}_I$ and $\mathcal{Q}_I$ are white noise parameters for pulsar $I$ (usually denoted as EFAC and EQUAD, respectively), $\mathbb{I}$ is the identity matrix, and $\mathcal{P}_I(f)$ is a red noise power spectrum 
\be 
\mathcal{P}_I(f)=\mathcal{P}_I^{\rm int}(f)+\mathcal{P}_g(f)
\ee 
where
\be
\mathcal{P}_I^{\rm int}(f)=\frac{A_{I}^2}{12\pi^2}
\left(\frac{f}{f_{\rm 1yr}}\right)^{2\alpha_I}f^{-3}
\ee
is the intrinsic red noise in the pulsar parameterized by amplitude $A_I$  and spectral index $\alpha_I$, and 
\be
\mathcal{P}_g(f)=\frac{A_{\rm gw}^2}{12\pi^2}
\left(\frac{f}{f_{\rm 1yr}}\right)^{2\alpha}f^{-3}
\label{powg}
\ee
is the gravitational-wave background spectrum parameterized by the
strain amplitude $A_{\rm gw}$ and spectral index $\alpha$. In other words, the
auto-covariance matrix of the noise in pulsar $I$ consists of intrinsic white noise
parameterized by $\{\mathcal{F}_I,\mathcal{Q}_I\}$ and red noise
parameterized by $\{ A_I,\alpha_I,A_{\rm gw},\gamma \}$. Notice that
the gravitational-wave parameters do not have a pulsar label because
they are common to all pulsars.

Similarly, the  cross-covariance matrices are given by 
\be
X_{IJ}=\langle \mathbf{n}_I \mathbf{n}_J^T \rangle_{ij} 
= \chi_{IJ} \int_0^{\infty} df\> 
e^{2\pi i f \tau_{ij}} \mathcal{P}_g(f) 
\ee
where $\chi_{IJ}$ are the Hellings and Downs coefficients for pulsar pair $I,J$ defined in \Deqn{hdcurve}.

We now write the likelihood function for the timing residuals using Eqs. \ref{pre} and \ref{likey} as
\be
p(\delta\mathbf{t}|\vec\theta,\delta\boldsymbol{\xi})=\frac{\exp\lp-\frac{1}{2}(\delta\mathbf{t}-\mathbf{M}\delta\boldsymbol{\xi})^T\mathbf{\Sigma}_{n}^{-1}(\delta\mathbf{t}-\mathbf{M}\delta\boldsymbol{\xi})\rp}{\sqrt{\det(2 \pi \mathbf{\Sigma}_{n})}},
\ee
where $\delta\mathbf{t}$ and $\delta\boldsymbol{\xi}$ are defined in
an identical manner as $\mathbf{n}$ as the concatenated vector or
residuals and timing parameters for each pulsar, respectively. Note
that here we use $\delta\mathbf{t}$ instead of $\delta\mathbf{t}^{\rm pre}$
since this process can be thought of as another step in the iterative
process of timing (where the post-fit residuals are formed from the
previous set of pre-fit residuals); instead of minimizing
chi-squared using $\mathbf{W}$ as the noise covariance, we now use the
full noise covariance matrix $\boldsymbol{\Sigma}_n$ and the full PTA
dataset to \emph{maximize} the likelihood. In \cite{vhl12} it was
shown that
this likelihood can be maximized\footnote{In \cite{vhl12}, the
  authors actually
  \emph{marginalize} the likelihood function over the pulsar timing
  parameters; however, when using uniform priors the resulting
  likelihood after maximizing or marginalizing only differs by a
  factor of $\det(\mathbf{M}^T\boldsymbol{\Sigma}_n\mathbf{M})$, so
  the data dependent part of the likelihood remains the same.}
analytically over the timing model parameters to give
\be
p(\delta\mathbf{t}|\vec\theta) = \frac{\exp\lp -\frac{1}{2}\delta\mathbf{t}^T\mathbf{G}(\mathbf{G}^T\boldsymbol{\Sigma}_n \mathbf{G})^{-1}\mathbf{G}^T\delta\mathbf{t} \rp}{\sqrt{\det(2\pi \boldsymbol{\Sigma}_n)}},
\ee
where $\mathbf{G}_I$ is an $N_{\rm TOA}\times(N_{\rm TOA}-N_{\rm par})$ matrix. The matrix $\mathbf{G}_I^T$ spans the null space of $\mathbf{M}_I$ and will project the data onto a subspace orthogonal to the linearized timing model. The full PTA G-matrix is then
\be
\label{gmat}
\mathbf{G}=\bb  \mathbf{G}_{1} & 0 & \hdots & 0\\ 
0 & \mathbf{G}_{2} & \hdots & 0\\
\vdots & \vdots & \ddots & \vdots\\
0 & 0 & \hdots & \mathbf{G}_{M}\eb.
\ee
For the remainder of paper we will use the following notation
\begin{align}
\mathbf{r}_I &= \mathbf{G}_I^T\delta\mathbf{t}_I\\
\mathbf{P}_I  &= \mathbf{G}_I^T\mathbf{N}_I\mathbf{G}_I\\
\mathbf{S}_{IJ} &= \mathbf{G}_I^T\mathbf{X}_{IJ}\mathbf{G}_J\\
\boldsymbol{\Sigma} &= \mathbf{G}^T\boldsymbol{\Sigma}_n\mathbf{G},
\end{align}
with the  likelihood function written as
\be
\label{liker}
p(\mathbf{r}|\vec\theta)=\frac{1}{\sqrt{\det(2 \pi \mathbf{\Sigma}_n)}}\exp\lp-\frac{1}{2}\mathbf{r}^T\mathbf{\Sigma}^{-1}\mathbf{r}\rp.
\ee

\subsubsection{Time-Domain Optimal Statistic}

In~\cite{abc+09} some of us presented the optimal cross-correlation
statistic in both the frequency and time domains, with a focus on the
frequency-domain implementation. The non-stationarity that arises from
the timing model fit (\Deqn{Rmatrices}), along with the irregular
sampling that is typical of realistic PTA data sets, however, make
frequency-domain techniques unsuitable for PTA gravitational-wave data
analysis. Therefore in this paper we will focus on the time-domain
implementation of the cross-correlation statistic. In \cite{abc+09}
the time-domain derivation was done by constructing the likelihood
ratio of a model that contained a stochastic gravitational-wave
background and intrinsic noise to a model that contained only
intrinsic noise. It was assumed that the amplitude of the intrinsic
noise is much larger than the amplitude of the gravitational-wave
background, and thus can be safely ignored in the auto-covariance
matrices of the residuals. One can then perform an expansion of the
log-likelihood ratio in powers of a small order parameter taken to
represent the amplitude of the background.  This assumption can lead
to a significant bias in the recovered amplitude of the
gravitational-wave background if the background is sufficiently large.

Fortunately it is possible to carry out a nearly identical derivation
that takes into account a potential non-negligible contribution of the
stochastic background to the auto-covariance terms.  In \cite{esvh13}
it was shown that it is possible to expand the covariance matrix
$\boldsymbol\Sigma$ in a Taylor series expansion in the Hellings and
Downs coefficients (as opposed to an expansion in the amplitude of
the background) to obtain a ``first order''
likelihood function. The log of this likelihood function can be
written as
\begin{multline}
\ln p(\mathbf{r}|\vec\theta)\approx-\frac{1}{2}\bigg[\sum_{I=1}^M \left(\tr\ln \mathbf{P}_{I}+\mathbf{r}_I^T\mathbf{P}_{I}^{-1}\mathbf{r}_I\right)\\
-\sum_{IJ}\mathbf{r}_I^T
\mathbf{P}_{J}^{-1}\mathbf{S}_{IJ}\mathbf{P}_{J}^{-1}\mathbf{r}_{J}\bigg]
 \label{foLike} 
 \end{multline} 
where
$ \sum_{IJ} = \sum_{I=1}^M\sum_{J<I}^M$
is a sum over all \emph{unique} pulsar pairs.
Let us now assume that we have done a single pulsar noise analysis
\cite{vhl12,esd+13} on each pulsar so that we know 
$\mathbf{P}_I$, and consider the following log-likelihood ratio
\be
\label{likeratio}
\ln\,\Lambda = \ln p(\mathbf{r}|\vec\theta_{\rm gw}) - \ln p(\mathbf{r}|\vec\theta_{\rm noise}).
\ee
Here $\vec\theta_{\rm gw}$ are the parameters for a model with 
a \emph{spatially correlated}\footnote{By spatially correlated 
we mean that the correlation is parameterized
by the Hellings and Downs curve.} 
gravitational-wave background component along with uncorrelated red and white noise components,
which include the gravitational-wave background 
present in the pulsar term, ISM noise, radiometer noise, jitter noise, etc.
The parameters $\vec\theta_{\rm noise}$ are for a model with 
only \emph{spatially uncorrelated} noise
components. We treat the
auto-covariance of each pulsar as a known measured quantity of the PTA
data after the aforementioned noise analysis has been done. In this
case, if we fix the spectral index to, say, the one corresponding to
SMBBH backgrounds with a spectral index $\alpha=-2/3$, the only free parameter is the amplitude of the gravitational-wave background. 
Evaluating this log-likelihood ratio we have
\be
\label{llrat}
\ln\,\Lambda = \frac{A_{\rm gw}^2}{2}\sum_{IJ}\mathbf{r}_I^T \mathbf{P}_{J}^{-1}\tilde{\mathbf{S}}_{IJ}\mathbf{P}_{J}^{-1}\mathbf{r}_{J},
\ee
where we have used the amplitude-independent cross-correlation matrix
$\tilde{\mathbf{S}}_{IJ}$ defined by
\be
\label{stilde}
A_{\rm gw}^2 \tilde{\mathbf{S}}_{IJ}= \langle \mathbf{r}_I \mathbf{r}_J^T\rangle = \mathbf{S}_{IJ}.
\ee
Notice that all terms that only include the auto-covariance matrices are 
cancelled by the noise model likelihood function. 
Note also that this expression is nearly identical to  Eq.~(75) of \cite{abc+09} 
with the caveat that now we are dealing exclusively with post-fit quantities and  
have allowed for a non-negligible contribution from the gravitational-wave background  
in the auto-covariance matrices. From Eq.~(\ref{llrat}) 
we define the optimal cross-correlation statistic for a PTA to be
\be
\label{os}
\hat A^2 =\frac{\sum_{IJ}\mathbf{r}_I^T \mathbf{P}_I^{-1}\tilde{\mathbf{S}}_{IJ}\mathbf{P}_J^{-1} \mathbf{r}_J}
{ \sum_{IJ} {\tr} \left[ \mathbf{P}_I^{-1}\tilde{\mathbf{S}}_{IJ}\mathbf{P}_J^{-1}\tilde{\mathbf{S}}_{JI} \right]},
\ee
where the normalization factor  
\be
{\mathcal N}\equiv
\left(\sum_{IJ}{\tr}\left[\mathbf{P}_I^{-1}\tilde{\mathbf{S}}_{IJ}\mathbf{P}_J^{-1}\tilde{\mathbf{S}}_{JI} 
\right]\right)^{-1}
\ee
is chosen so that on average  $\langle \hat A^2\rangle=A_{\rm gw}^2$. 
This immediately follows from the observation that
\be
\begin{aligned}
\Big\langle\sum_{IJ}\mathbf{r}_I^T \mathbf{P}_I^{-1}\tilde{\mathbf{S}}_{IJ}\mathbf{P}_J^{-1} \mathbf{r}_J\Big\rangle &=
 \sum_{IJ} {\tr} \left[ \mathbf{P}_I^{-1}\tilde{\mathbf{S}}_{IJ}\mathbf{P}_J^{-1}\mathbf{S}_{JI} \right]\\
 &=A_{\rm gw}^2\sum_{IJ} {\tr} \left[ \mathbf{P}_I^{-1}\tilde{\mathbf{S}}_{IJ}\mathbf{P}_J^{-1}\tilde{\mathbf{S}}_{JI} \right],
\end{aligned}
\ee
where Eq.~(\ref{stilde}) was used in the second line.

In the absence of a cross-correlated signal (or if the signal is weak)
the expectation value of $\hat A^2$ vanishes and its
standard deviation is~\cite{abc+09}
\be
\label{std}
\sigma_0 =\left(\sum_{IJ} \tr \left[ \mathbf{P}_I^{-1}\tilde{\mathbf{S}}_{IJ}\mathbf{P}_J^{-1}\tilde{\mathbf{S}}_{JI} \right]\right)^{-1/2},
\ee
so if in a particular realization we measure a value of
the optimal statistic, the signal-to-noise ratio (SNR) 
for the \emph{power in the cross-correlations} for that realization is
\be
\hat{\rho}=\frac{\hat A^2}{\sigma_0}=\frac{\sum_{IJ}\mathbf{r}^T_I \mathbf{P}_I^{-1}\tilde{\mathbf{S}}_{IJ}\mathbf{P}_J^{-1} \mathbf{r}_J}
{ \left(\sum_{IJ} \tr \left[ \mathbf{P}_I^{-1}\tilde{\mathbf{S}}_{IJ}\mathbf{P}_J^{-1}\tilde{\mathbf{S}}_{JI}
  \right] \right)^{1/2} }.
\label{SNR}
\ee
with an expectation value over all realizations of 
\be
\langle \rho \rangle = A_{\rm gw}^2  \left( \sum_{IJ} \tr
  \left[ \mathbf{P}_I^{-1}\tilde{\mathbf{S}}_{IJ}\mathbf{P}_J^{-1}\tilde{\mathbf{S}}_{JI} \right] \right)^{1/2}.
\label{rhoav}
\ee 
Note that this definition of the SNR measures the confidence (in
standard deviations) with which we can reject the null hypothesis that
there are no \emph{spatially correlated} signals in our data.  To
clarify this a bit further we outline a standard frequentist
hypothesis detection procedure:
\begin{enumerate}
 \item Measure the optimal statistic value, $\hat{A}^2$ of Eq.~(\ref{os}), for our data set.
 
 \item Compute the probability $p(\hat{A}^2>\hat{A}_{\rm thresh}^2 |
   A_{\rm gw}=0)$, that is, the probability that our measurement of
   the optimal statistic, $\hat{A}^2$, is greater than some threshold
   value of the statistic, $\hat{A}_{\rm thresh}^2$, assuming that the
   null hypothesis, $A_{\rm gw}=0$, is true.
 
 \item If the aforementioned probability (sometimes called the
   $p$-value) is less than some value (this value is set to be a
   tolerable yet problem specific false-alarm probability (FAP)) then
   a detection is claimed.
 
 \end{enumerate}
 Typically $\hat{A}_{\rm thresh}^2$ is given by
 \be
 \alpha = \int_{-\infty}^{\hat{A}_{\rm thresh}^2} d\hat{A}^2 p(\hat{A}^2|A_{\rm gw}=0),
 \ee
 where $\alpha$ is the FAP and $p(\hat{A}^2|A_{\rm gw}=0)$ is the
 probability distribution function of the optimal statistic given the
 null hypothesis. To a sufficiently good approximation,
 $p(\hat{A}^2|A_{\rm gw}=0)$ can be described by a Gaussian
 distribution with zero mean and variance given by $\sigma_0^2$
 (Eq.~(\ref{std})), thus the probability $p(\hat{A}^2>\hat{A}_{\rm
   thresh}^2 | A_{\rm gw}=0)$ can be expressed in terms of standard
 deviations away from the mean. For example, if the $\hat{A}^2$ that
 we measure is 3 standard deviations (i.e 3-sigma) away from the mean
 (0 in this case) then this corresponds to a FAP of $\sim$0.003
 meaning that we can rule out the null hypothesis with  $\sim$99.7\%
 confidence. Returning to Eq.~(\ref{SNR}) we see that the typical
 frequentist detection procedure mentioned above is
 contained in this definition of SNR. If we measure an SNR of 3, this
 carries the same meaning as the FAP above.
 
\begin{figure}
\includegraphics[width=3.4in]{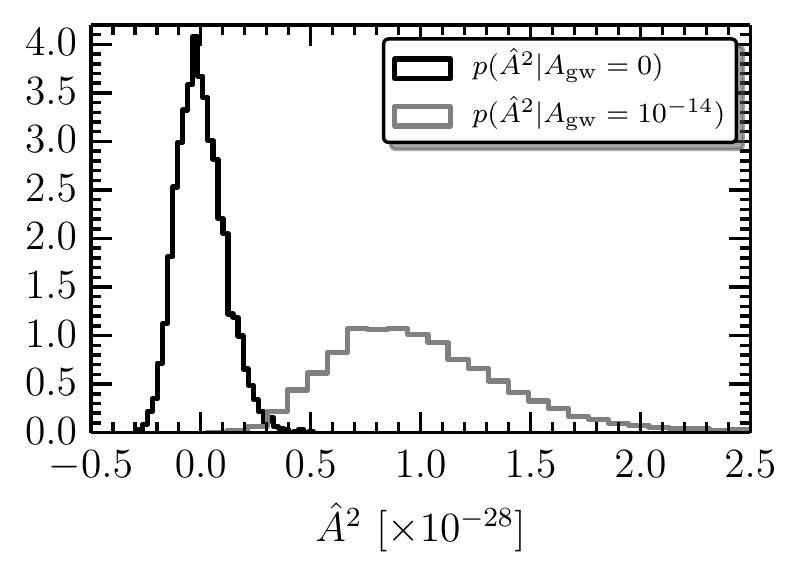}
\caption{Histogram of the optimal
statistic Eq.~(\ref{os}) in $10^4$ simulations for a PTA consisting of
$M=36$ pulsars, all with
RMSs $\sigma=100$~ns, an
observational time $T=5$ years, and a cadence $c=20$~yr$^{-1}$. We show
the distribution of the statistic in the absence of a signal (black line), and the
the distribution in the presence of
a signal with amplitude $A_{\rm gw}=10^{-14}$ (gray line). The standard
deviation of the distribution in the absence of a signal is $\sigma_0 = 1.08 \times 10^{-29}$.}
\label{fig:opt_stat_hist_bw}
\end{figure}

Figure~\ref{fig:opt_stat_hist_bw} shows a histogram of the optimal
statistic Eq.~(\ref{os}) in $10^4$ simulations for PTA observations
of $M=36$ pulsars, with root-mean-squares (RMSs) $\sigma=100$~ns, for an observational
time $T=5$ years, and a cadence $c=20$~yr$^{-1}$. The black line shows
the distribution of the statistic in the absence of a signal, and the
gray curve shows the distribution in the presence of a signal with
amplitude $A_{\rm gw}=10^{-14}$ (using the methods described below in
Section~\ref{sec:inj}). The standard deviation of the distribution in
the absence of a signal is $\sigma_0 = 1.08 \times 10^{-29}$.  As
shown in the figure, in the absence of a signal the distribution is
not quite Gaussian, but using the true cumulative distribution of the
simulations and the 3-$\sigma$ Gaussian distribution threshold gives a
FAP of $\sim 0.006$.

\section{Simulated signals}
\label{sec:inj}

In this section we describe a software injection procedure that can be
used to produce simulated stochastic background 
signals in PTA data.  As we have shown, if a stochastic gravitational-wave 
background is present, the cross-correlation of timing residuals is given by
\be
\langle \tilde{r}^*_I(f) \tilde{r}_J(f') \rangle = \frac{H_0^2}{16 \pi^4} \delta(f-f')  |f|^{-5}  \Omega_{\rm gw}(f) \chi_{IJ}. 
\label{expcorr}
\ee
In the frequency domain it is possible to express the 
timing residuals as
\be 
r_I(f) = c(f) \sum_{J} H_{IJ} w_J(f), 
\label{rasw}
\ee
where  $w_I(f) = x_I(f) + i y_I(f)$ is a complex zero-mean white noise
process, $c(f)$ is a real function that contains information about the spectral
index and amplitude of the gravitational-wave spectrum 
(but does not depend on the pulsar pair), and $H_{IJ}$ is a matrix that 
linearly combines the timing residuals in
such a way as to simulate the expected spatial correlations in the
signal, i.e. the Hellings and Downs coefficients.

If the processes $x_I$ and $y_I$ are zero-mean unit-variance
processes $w_I(f)$ satisfies 
\be
\langle w^*_I(f) w_J(f') \rangle = \frac{2}{T} \delta(f-f') \delta_{IJ},
\ee
where $T$ is the length of observations, and we can 
use Eq.~(\ref{expcorr}) to find $c(f)$ and $H_{IJ}$.  Taking the ensemble average of \Deqn{rasw} it is easy to show that
\be
\langle \tilde r^*_I(f) \tilde{{r}_J}(f') \rangle =\frac{2}{T} c(f) c(f') H_{IJ} H_{JI} \delta(f-f'),
\label{nexpcor}
\ee
which implies that
\be
 c^2(f) H_{IJ} H_{JI} = \frac{T H_0^2}{32\pi^4}|f|^{-5} \Omega_{\rm gw}(f)  \chi_{IJ}.
\ee
In matrix notation the equation above can be written as 
\be 
c^2(f) {\bf H H^T} = \frac{T H_0^2}{32\pi^4}|f|^{-5} \Omega_{\rm gw}(f) {\boldsymbol \chi}.
\label{matrixeqn}
\ee
Relating the functions of frequency on either side of
\Deqn{matrixeqn}, we readily identify the function $c(f)$ to be
\be
c(f) = \left[ \frac{T H_0^2}{32\pi^4}\Omega_{\rm gw}(f) |f|^{-5} \right]^{1/2},
\ee
along with a condition for the matrix ${\bf H}$,
\be
{\bf HH^T} = {\boldsymbol \chi}
\ee
which allows us to determine $\bf H$ given $\boldsymbol
\chi$ via a Cholesky decomposition.

To construct simulated timing residuals one can:
(1) start with $M$ random complex frequency series 
$w_I(f)$, where $M$ is the number of pulsars,
(2) multiply these by $c(f)$, 
(3) find the Hellings and Downs coefficients for all pulsar
pairs and construct the matrix ${\boldsymbol \chi}$, 
(4) perform a Cholesky decomposition of ${\boldsymbol \chi}$ 
to find ${\bf H}$, and (5) linearly combine the frequency
series via Eq.~(\ref{rasw}) to find $r_I(f)$ for each pulsar. 
Finally, after inverse Fourier transforming the 
gravitational-wave residuals, they
can be added to real or simulated TOA
data that contains additional uncorrelated white and red noise components.

\section{Scaling Laws for the Optimal Cross-correlation statistic}
\label{sec:scalingLaws}

In~\cite{sej+13} the authors considered a simple scenario where pulsar
timing residuals have just two noise components, a
gravitational-wave red noise piece and a white-noise piece, which
are the same for all pulsars in the PTA, namely
\be
\mathcal{P}_I(f)=\mathcal{P}_g(f)+ 2\sigma^2\Delta t 
= bf^{-\gamma} + 2\sigma^2\Delta t.
\label{ps}
\ee
Here all the frequency independent
constants in Eq.~(\ref{powg}) have been absorbed
into the amplitude $b$, the index $\gamma = 3-2\alpha$
(recall that we are using one-sided power spectra
in this paper, in contrast to \cite{sej+13}), and
the white noise RMS is denoted by $\sigma$.  

In~\cite{sej+13} it was shown that the SNR of the optimal
cross-correlation scales in three different ways depending on the
relative sizes of the gravitational-wave and white-noise
components. Specifically the authors found scaling laws for the SNR in  
\begin{enumerate}[(i)]
	\item a weak signal regime where the white noise
          component of \Deqn{ps} is larger than the gravitational wave
          piece ($2\sigma_I^2\Delta t \gg bf^{-\gamma} $ at all
          relevant frequencies),
	\item the opposite strong signal limit, where 
          $2\sigma_I^2\Delta t \ll bf^{-\gamma}$ at all relevant frequencies, which turns out to be
          irrelevant for pulsar timing experiments, and,  
	\item an intermediate regime between the two cases where the
          gravitational wave power spectrum dominates at low
          frequencies, and the white noise dominates at high frequencies.
\end{enumerate}
Additionally, they found that the latter regime is likely already
relevant to current pulsar timing experiments. In this section we will
review the scaling laws for the optimal statistic, and introduce an
improved derivation of the scaling law for the intermediate regime.

To derive the scaling laws we begin with the expression for the
expected SNR of the cross-correlation statistic,
\be
\langle \rho \rangle = A_{\rm gw}^2  \left( \sum_{IJ} \tr
  \left[ \mathbf{P}_I^{-1}\tilde{\mathbf{S}}_{IJ}\mathbf{P}_J^{-1}\tilde{\mathbf{S}}_{JI} \right] \right)^{1/2},
\label{rhoav2}
\ee
which can be written in the frequency domain as~\cite{abc+09}
\be 
\langle \rho \rangle = \left(2 T \sum_{IJ} \chi_{IJ}^2
  \int_{f_L}^{f_H} df\, 
\frac{{\mathcal P}_g^2(f)}{{\mathcal P}_I(f) {\mathcal P}_J(f)} \right)^{1/2}.
\label{intSNR}
\ee
Since we are assuming that all pulsars have the same noise
characteristics we can write
\be 
\langle \rho \rangle = \left(\sum_{IJ} \chi_{IJ}^2 \right)^{1/2}
\left(2T \int_{f_L}^{f_H} df\,\frac{b^2
    f^{-2\gamma}}{\left(bf^{-\gamma}+2\sigma^2 \Delta t \right)^2} \right)^{1/2}.
\label{SNRallsame}
\ee

In the weak signal regime, where $2\sigma_I^2\Delta t \gg bf^{-\gamma}
$ for all frequencies of interest, i.e.,  
$f \in [f_L,f_H]$, the SNR is well approximated by  
\be 
\langle \rho \rangle \approx \left(\sum_{IJ} \chi_{IJ}^2 \right)^{1/2}
\frac{b c T^\gamma}{2\sigma^2 \sqrt{\gamma-1/2}},
\label{weaksignal}
\ee
where $c=1/\Delta t$ is the cadence.

In the intermediate regime we cannot use this approximation because at
low frequencies the power in the gravitational-wave backround is larger
than the white noise level. Note that this happens when
$bT^\gamma >  2\sigma^2\Delta t$, and the condition on
the white noise RMS is
\be
\sigma < \frac{A}{\pi f_{\rm 1yr}^\alpha}\sqrt{\frac{cT^\gamma}{24}}.
\label{RMSthreshold}
\ee
For pulsar timing experiment durations of $T=5$~yr,
cadence of $c=20$~yr$^{-1}$, for a background with amplitude
$A=10^{-15}$, and a spectral index like the one we expect for the SMBBH
background ($\gamma=13/3$), the pulsar timing array is in the weak signal limit only
if the pulsars have white noise RMSs greater than about
$300$~ns. There are already a handful of pulsars that are currently
timed with better precisions than that (see, for example,~\cite{dfg+12}).

In this case the integral in \Deqn{SNRallsame} must evaluated
generally. To do this, we write the integral as
\be 
\int_{f_L}^{f_H} df\, F(f) = \int_0^{f_H} df\, F(f) -\int_0^{f_L} df\, F(f)
\label{sigintsum}
\ee
where for convenience we have written 
\be 
F(f) = \frac{ b f^{-2\gamma} }{\left(b f^{-\gamma}+2\sigma^2 \Delta t \right)^2}.
\ee
The integrals on the right hand side of Eq.~(\ref{sigintsum}) have analytic solutions in terms of of ordinary hypergeometric functions. 
To proceed, we evaluate the integral of $F(f)$ over a generic interval $[0, f_*]$ which yields 
\be
%\begin{split}
\int_0^{f_*} df F(f) = \frac{f_*}{\gamma}
\left[\frac{1}{1+\frac{2\sigma^2 \Delta t}{b f_*^{-\gamma}}}
+ \left( \gamma -1\right) G\left(\frac{-2\sigma^2 \Delta t}{b f_*^{-\gamma}}\right)\right],
%\end{split}
\label{genint}
\ee
where $G(x)=\,_2F_1\left(1,\gamma^{-1},1+\gamma^{-1},x\right)$. We can probe this
solution in the context of Eq.~(\ref{sigintsum}) by replacing $f_*$ with $f_H$ or $f_L$. 

For the second integral on the right hand side of Eq.~(\ref{sigintsum}) where $f_* = f_L = 1/T$, we have $(2\sigma^2 \Delta t)/(b
f_L^{-\gamma}) \ll 1$ and the hypergeometric function can be
approximated to be unity 
$$
_2F_1\left(1,\gamma^{-1},1+\gamma^{-1},\frac{-2\sigma^2 \Delta t}{b
    f_L^{-\gamma}}\right) \approx 1.$$ 
This simplifies \Deqn{genint}
greatly, and the integral is easily evaluated as
\be \int_0^{f_L} df\,
F(f) \approx \frac{1}{T}.
\label{fLsoln}
\ee

To evaluate the first integral in Eq.~(\ref{sigintsum}), we consider the case when $f_* = f_H$
in Eq.~(\ref{genint}). 
In this case, since $(2\sigma^2 \Delta t)/(b
f_H^{-\gamma}) \gg 1$, the integral can be approximated as
\be
%\begin{split}
\int_0^{f_H} df\, F(f) \approx \frac{f_H}{\gamma}
\left[\frac {b f_H^{-\gamma}}{2\sigma^2 \Delta t}
+ \left( \gamma -1\right) G\left(\frac{-2\sigma^2 \Delta t}{b f_H^{-\gamma}}\right)\right],
%\end{split}
\label{genint2}
\ee
We can then use 
standard identities relating the hypergeometric function to inverses
of their arguments (see,
for example, Eq.~(15.8.2) in \cite{olb+10}). Using these identities
along with with Euler's reflection formula we obtain 
\begin{widetext}
\begin{multline}
\int_0^{f_H} df\, F(f) \approx  
\frac{f_H}{\gamma} 
\left\{ \frac{b f_H^{-\gamma}}{2\sigma^2 \Delta t}+ \left( \gamma -1\right)  \, \Gamma(\gamma^{-1}-1) \Gamma(2-\gamma^{-1}) \Gamma(1+\gamma^{-1}) \right. 
\\
\left. \times \left[ \frac{b f_H^{-\gamma}}{2\sigma^2 \Delta t}
    \frac{\Gamma(2-\gamma^{-1})^{-1}}{\Gamma(\gamma^{-1})^2} 
\, _2F_1\left(1,1-\gamma^{-1},2-\gamma^{-1};  \frac{-b f_H^{-\gamma}}{2\sigma^2 \Delta t} \right)
- \frac{1}{\Gamma(\gamma^{-1})}\left(\frac{b f_H^{-\gamma}}{2\sigma^2 \Delta t}  \right)^{1/\gamma} 
\,_2F_1\left(\gamma^{-1},0,\gamma^{-1}; \frac{-b f_H^{-\gamma}}{2\sigma^2 \Delta t}  \right) \right] \right\}.
\label{expanded_fHcase}
\end{multline}
\end{widetext}
Since $b f_H^{-\gamma}/2\sigma^2 \Delta t \ll 1$ both hypergeometric functions can be well
approximated by unity. Additionally,
since $b f_H^{-\gamma}/2\sigma^2 \Delta t \ll \left(b f_H^{-\gamma}/2\sigma^2 \Delta t\right)^{1/\gamma}$
for $\gamma > 1$, the last term in \Deqn{expanded_fHcase}
dominates and the expression can be simplified to
\be
\int_0^{f_H} df\, F(f) \approx \kappa(\gamma) \left(\frac{b}{2\sigma^2 \Delta t} \right)^{1/\gamma}
\label{fHsoln}
\ee
with 
\be
\kappa(\gamma)= \frac{\left(1- \gamma \right) \Gamma(\gamma^{-1}-1)\Gamma(2-\gamma^{-1})\Gamma(1+\gamma^{-1})}{\gamma \,\Gamma(\gamma^{-1})}.
\ee

Putting the results of \Deqn{fLsoln} and \Deqn{fHsoln} together, we arrive at the solution to the original problem posed 
in \Deqn{sigintsum}:
\be
\int_{f_L}^{f_H} df \, F(f) \approx \kappa(\gamma) \left( \frac{b}{2\sigma^2 \Delta t} \right)^{(1/\gamma)} - \frac{1}{T}.
\ee
In terms of the cadence $c=1/\Delta t$ the average value of the SNR is therefore given by
 \be
\langle \rho \rangle \approx \left(\sum_{IJ} \chi^2_{IJ}\right)^{1/2}
\left[ 2T \left( \kappa(\gamma) \left( \frac{bc}{2\sigma^2}
    \right)^{(1/\gamma)} - \frac{1}{T} \right)  \right]^{1/2}.
\label{intermediate}
\ee
At late times, 
\be
\begin{aligned}
\label{intermediatelatetimes}
\langle \rho \rangle &\approx 
\left(\sum_{IJ} \chi^2_{IJ}\right)^{1/2} 
\left[ 2T \kappa(\gamma) \left( \frac{bc}{2\sigma^2 } \right)^{(1/\gamma)} \right]^{1/2}  
\\
&\propto M \left( \frac{c A_{\rm gw}^2}{2\sigma^2} \right)^{1/(2\gamma)} T^{1/2}.
\end{aligned}
\ee

In~\cite{sej+13} the authors approximated the integral in a less accurate (albeit
more pedagogical) way: they found the frequency $f_r=(bc/2\sigma^2)^{1/\gamma}$ at which the
gravitational wave red noise equals the white noise, and assumed
the integral was gravitational wave dominated at frequencies lower
than $f_r$, and white noise dominated at frequencies higher than
$f_r$. The integrals then become trivial. The result is the same as Eq.~(\ref{intermediate}), but with a
different value of the coefficient $\kappa$ which was found to
be $\kappa'=2\gamma/(2\gamma-1)$. In the approximation the integrand
for the SNR is always over-estimated and 
the value of $\kappa'$ is larger than what we have
calculated for $\kappa$ in this paper.

\begin{figure}
%\centering
\includegraphics[width=3.5in]{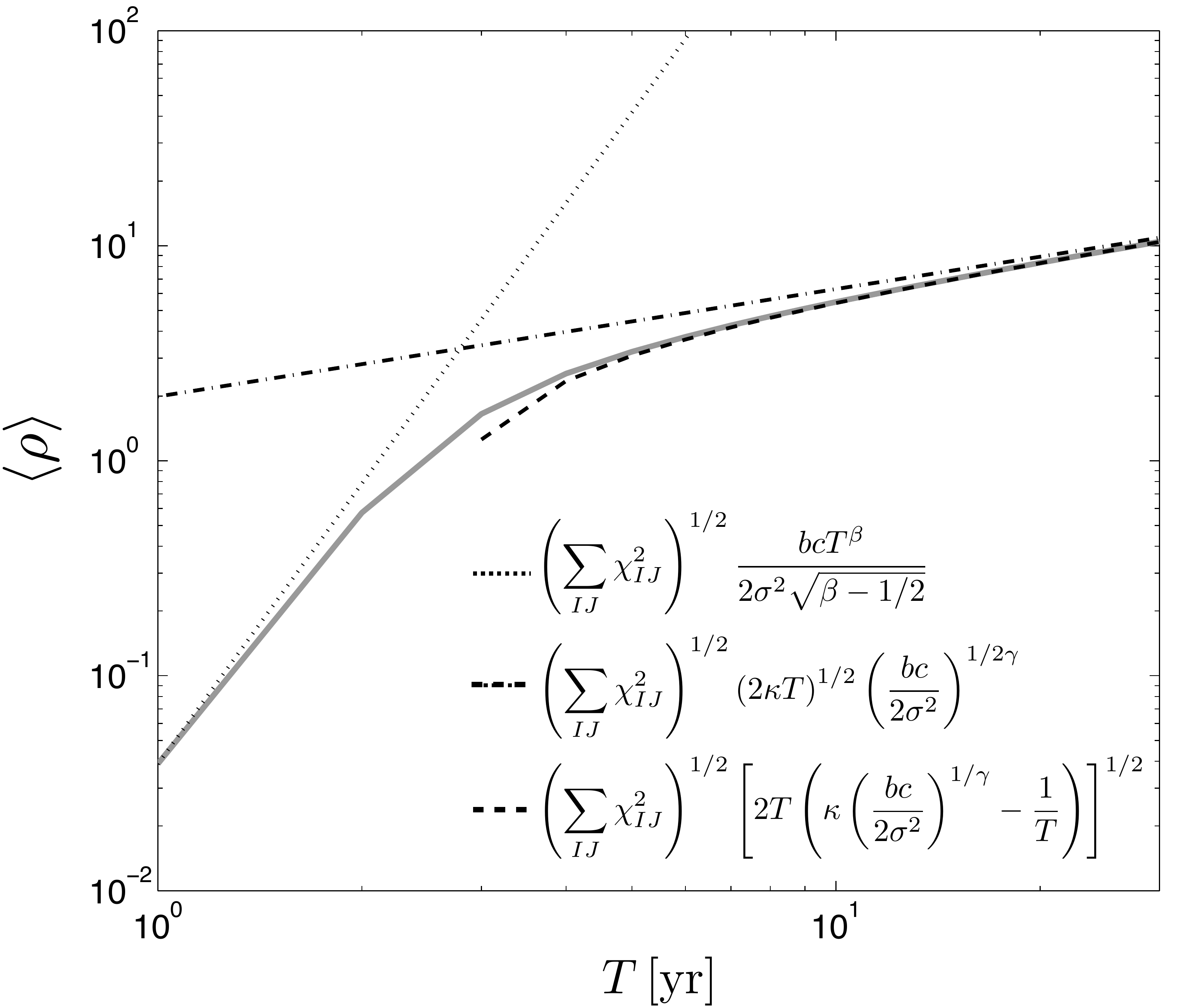}
\caption{Average SNR versus time in years for PTA
with 20 pulsars timed with a presicion of $\sigma=50$~ns and a
gravitational-wave background produced by SMBBHs ($\gamma=13/3$) with
an amplitude $A_{\rm gw}=10^{-15}$. The gray curve shows the SNR computed
numerically using Eq.~(\ref{rhoav2}). 
The dotted curve shows  SNR in the weak-signal
limit, Eq.~(\ref{weaksignal}). The dashed-dot curve shows the
SNR in the intermediate regime at late times,
Eq.~(\ref{intermediatelatetimes}). The dashed curve shows the
SNR calculated using Eq.~(\ref{intermediate}).
}\label{fig:numAgree}
\end{figure}

Figure~\ref{fig:numAgree} shows the average SNR versus time in years for PTA
with 20 pulsars timed with a presicion of $\sigma=50$~ns and a
gravitational-wave background produced by SMBBHs ($\gamma=13/3$) with
an amplitude $A_{\rm gw}=10^{-15}$. The gray curve shows the SNR computed
numerically in the time domain using Eq.~(\ref{rhoav2}). For the
timing model we have subtracted out a quadratic---i.e., we have fitted
out a quadratic with the $\mathbf{R}$ projection matrices in the time domain.
The dotted curve shows the average SNR as computed in the weak-signal
limit using Eq.~(\ref{weaksignal}). The dashed-dot curve shows the
SNR in the intermediate regime at late times as calculated using 
Eq.~(\ref{intermediatelatetimes}). Finally, the dashed curve shows the
SNR calculated using Eq.~(\ref{intermediate}). 
At very early times the approximation is not valid: the first term in
the square root is smaller than $1/T$ so the SNR is imaginary. At
later times the predicted SNR is in excellent agreement with the
time-domain numerical calculation. Note the remarkable accuracy with
which the low frequency cutoff $f_L=1/T$ approximates the effect of
quadratic subtraction.

\section{Summary}
\label{sec:conclude}

In this paper, we have presented a time-domain implementation of the
optimal cross-correlation statistic for stochastic gravitational-wave
background searches using PTA data, originally presented in
\cite{abc+09}.  The derivation and implementation described here
extends that of \cite{abc+09} by taking the timing model into account
in a natural and statistically well-motivated way by including the
linear timing model directly into the likelihood function, allowing
for analytic maximization of the timing model parameters.  The
time-domain implementation also allows one to fully model the noise
and naturally deal with non-stationarities and irregular sampling of
the data, which cannot be modeled in the frequency domain.

An alternative approach for analyzing PTA data for stochastic
gravitational-wave backgrounds is to use Bayesian inference, as
described in \citep{vlm+09, vhj+11, esvh13, lah12,
  tgl12}.  In the Bayesian approach, one constructs the
posterior probability distributions for the noise and
gravitational-wave signal parameters via Bayes' theorem by specifying
the likelihood function for the data given a set of model 
gravitational-wave and noise
parameters and a prior distribution on the model parameters.  By
marginalizing over the model parameters, one also constructs the
Bayesian evidence for various models, which allow for the construction
of Bayes factors (ratio of Bayesian evidence) to determine which model
is favored by the data.

While we believe that a Bayesian approach to the detection problem for
stochastic backgrounds is preferred and indeed
recommended, the frequentist cross-correlation statistic presented
here has several advantages over the Bayesian approach. Firstly, the
optimal statistic approach is computationally inexpensive as it
involves only a single function call (given a set of modeled noise
parameters), while the Bayesian method must explore a very large
dimensional space leading to millions of likelihood evaluations. For
current data sets, the optimal statistic can be evaluated in seconds
while the full Bayesian approach can take weeks to run on a super
computer.

Furthermore, the SNR as defined in this work is a good approximation
to the Bayes factor comparing a model for a correlated
gravitational-wave background to a model for an uncorrelated intrinsic
red noise source.  Thus the computationally inexpensive optimal
statistic has proven invaluable in large scale simulations and
projections of detector sensitivity as it allows us to test many
different signal models and pulsar observation scenarios with relative
ease, while full Bayesian simulations on this scale are unfeasible. In
addition, the relationship between the the optimal statistic SNR and
the Bayes factors affords an analytically tractable environment from
which to construct various scaling relations as shown in Figure
\ref{fig:numAgree}.

The optimal statistic does have two major drawbacks that make it less
desirable as a production-level detection statistic compared to the
Bayes factor.  Firstly, the point estimate of the amplitude of the
gravitational-wave background depends on our ability to
accurately model the total autocorrelated power for each
pulsar. Typically this is done by modeling the noise for each pulsar
\emph{independently} and then including the maximum likelihood values
in the auto-covariance matrices of the optimal statistic. If the
signal is loud and the data does not contain any intrinsic red noise
then this method is fairly robust and does not significantly bias
results.  However, if the signal is weak or there is other intrinsic
red noise then this method will lead to biases.  In low SNR scenarios
the red noise due to the stochastic background may not be large enough
to detect in an individual pulsar and will thus not enter the
auto-covariance matrices used in the optimal statistic. This will lead
to an inconsistency in the optimal statistic where it will still be
able to detect cross-correlated power, but the point estimate of the
amplitude will be biased low because the auto-covariance terms (from
our single pulsar noise analysis) indicate that the red noise is very
weak.

This problem does not arise in Bayesian analyses because the intrinsic
pulsar noise and the stochastic background parameters are modeled
\emph{simultaneously}.  This problem could be ameliorated by
performing the initial noise modeling over all pulsars simultaneously
and including a correlated gravitational-wave background
component. These noise estimates (which will include a common 
gravitational-wave
background term in the auto-covariance) could then be input to the
optimal statistic.

Despite these drawbacks, the optimal cross-correlation statistic
serves as a proxy for a full Bayesian search when performing
computationally-intensive simulations and will also serve as a very
useful cross-check when making detection statements on future PTA
data.

% SECTION: ACKNOWLEDGEMENTS
\begin{acknowledgments}

  We thank Eanna Flanagan for pointing out a sign error in~\cite{abc+09}, 
  which has been corrected here. We also thank Chris Pankow and Madeline Wade 
  for many useful comments and suggestions. We extend our gratitude to the members 
  of the NANOGrav Data Analysis Working Group. This work was funded in part by the 
  Wisconsin Space Grant Consortium and the National Science Foundation. 
  JDR would like to acknowledge support from NSF awards HRD-0734800 and HRD-1242090.
  JAE acknowledges support by NASA through Einstein Fellowship grant PF4-150120.

\end{acknowledgments}

%% Appendix section
\appendix

\section{Relation to Demorest et al. Cross-Correlation Statistic}

Here we show that the optimal statistic, although derived in a 
different manner, is identical to the cross-correlation statistic 
presented in \cite{dfg+12}. 
In the notation used in this work, the cross-correlation 
coefficients can be written as
\be
\label{rhoIJ}
\rho_{IJ} = \frac{\mathbf{r}_I^T\mathbf{P}_I^{-1}\hat{\mathbf{S}}_{IJ}\mathbf{P}_J^{-1}\mathbf{r}_J}{ {\tr} \left[ \mathbf{P}_I^{-1}\hat{\mathbf{S}}_{IJ}\mathbf{P}_J^{-1}\hat{\mathbf{S}}_{JI} \right]},
\ee
where $\hat{\mathbf{S}}_{IJ}$ is defined so that $A_{\rm gw}^2\chi_{IJ}\hat{\mathbf{S}}_{IJ} = \mathbf{S}_{IJ}$. 
The uncertainty on the correlation coefficients is 
\be
\label{sigmaIJ}
\sigma_{IJ} = \left(
\tr \left[ \mathbf{P}_I^{-1}\hat{\mathbf{S}}_{IJ}\mathbf{P}_J^{-1}\hat{\mathbf{S}}_{JI} \right]
\right)^{-1/2}.
\ee
With these expressions we now have an estimate of the cross-correlation coefficients along with their uncertainty for each pulsar pair. Notice that only the spectral shape of the gravitational wave background  is assumed. To determine an estimate of the gravitational wave background  amplitude, the following chi-squared is minimized
\be
\chi^2 = \sum_{IJ}\left(\frac{\rho_{IJ} - A_{\rm gw}^2\chi_{IJ}^2}{\sigma_{IJ}}\right)^2.
\ee
The resulting best fit gravitational wave amplitude is
\be
\hat{A}_{\rm gw}^2 = \sum_{IJ}\frac{\rho_{IJ}\chi_{IJ}}{\sigma_{IJ}^2}\bigg/ \sum_{IJ}\frac{\chi_{IJ}^2}{\sigma_{IJ}^2},
\ee
with variance
\be
\sigma^2 = \left( \sum_{IJ}\frac{\chi_{IJ}^2}{\sigma_{IJ}^2}\right)^{-1}.
\ee
By using Eqs. \ref{rhoIJ} and \ref{sigmaIJ} and by noting that $\chi_{IJ}\hat{\mathbf{S}}_{IJ} = \tilde{\mathbf{S}}_{IJ}$, we obtain
\be
\hat A_{\rm gw}^2 =\frac{\sum_{IJ}\mathbf{r}_I^T \mathbf{P}_I^{-1}\tilde{\mathbf{S}}_{IJ}\mathbf{P}_J^{-1} \mathbf{r}_J}
{ \sum_{IJ} {\tr} \left[ \mathbf{P}_I^{-1}\tilde{\mathbf{S}}_{IJ}\mathbf{P}_J^{-1}\tilde{\mathbf{S}}_{JI} \right]},
\ee
which is identical to Eq.~(\ref{os}).

\bibliographystyle{h-physrev}
\bibliography{apjjabb,bib}

\end{document}